\documentclass{KapProc} 
\usepackage[dvips]{graphicx}
\let\footnote\savefootnote
\let\footnotetext\savefootnotetext 
\let\footnoterule\savefootnoterule 
\normallatexbib

\begin{document}

\articletitle[Article Title at running head]
{Treatment of Pairing Correlations in Nuclei Close to Drip Lines}

\author[ Nguyen Van Giai, M. Grasso, R. Liotta and N. Sandulescu]%
{
 \ Nguyen Van Giai$^{\rm a}$,  M. Grasso$^{\rm b}$, R.J.
Liotta$^{\rm c}$ and N. Sandulescu$^{\rm d}$
}

\affil{$^{\rm a}$ Institut de Physique Nucl\'eaire, IN2P3-CNRS, \\ 
Universit\'e Paris-Sud,  91406 Orsay Cedex, France}

\affil{$^{\rm b}$ Universit\`a di Catania, 95129 Catania, Italy}

\affil{$^{\rm c}$ Royal Institute of Technology, Frescativ. 24, 10405
  Stockholm, Sweden}

\affil{$^{\rm d}$ NIPNE, POB MG-6, Bucharest-Magurele, Romania}




\begin{abstract}
We discuss the HFB equations in coordinate representation, a suitable method 
for handling the full effects of the continuous quasiparticle spectrum. 
We show how the continuum HFB equations can be solved with the correct 
asymptotic conditions instead of the discretization conditions which are 
commonly used in the literature. The continuum HFB method is illustrated 
with a model where the mean field and pairing field have simple  forms. 
The relationship with the continuum Hartree-Fock-BCS (HF-BCS) 
approximation is also discussed. 
Realistic HFB and HF-BCS calculations based on  Skyrme interactions are
compared for the case of a neutron-rich nucleus. 
 
\end{abstract}

\section{Introduction}

In many nuclear systems pairing correlations have an important influence
on physical properties. This situation occurs when the energy difference
between an occupied orbital and a neighbouring unoccupied one is relatively
small, thus enabling a nucleon pair to be promoted to the unoccupied level
by the interaction among the pair. Well known theoretical approaches such as
the Hartree-Fock-Bogoliubov (HFB) or the Hartree-Fock-BCS (HF-BCS)
approximations have been developed~\cite{Ring-Schuck} 
to treat the pairing effects in nuclei.

In usual situations, i.e., when the nuclear system is stable and far away 
from the drip lines the active orbitals are well bound and it may be sufficient 
to solve the HFB or HF-BCS equations within the discrete subspace of those 
active orbitals. When one approaches a drip line this is no longer true since 
the active orbitals must also include states belonging to the continuous 
single-particle spectrum. One of the most convenient tools for the
treatment of pairing correlations in the presence of continuum coupling
is given by the  HFB approach. The standard procedure is to discretize the 
quasiparticle continuum by solving the HFB equations in a finite basis of 
orthonormal functions, or by imposing  to the solutions of HFB equations,
written in coordinate space, a box boundary condition at some 
distance $R_b$ ~\cite{Doba84,Doba96,Tera96}. 
There may be circumstances where a discretized solution is 
unsatisfactory unless an extremely large box radius $R_b$ is used. 
Such should be the case for halo nuclei, for instance. In these lectures, 
we show how the HFB equations can be solved with correct boundary 
conditions. Based on these solutions we examine how much the continuum,
especially the resonant continuum, could affect the pairing properties
of nuclei close to the drip lines. 

The structure of the lectures is the following: In the next section we 
introduce the HFB equations in coordinate space ~\cite{Bulgac,Doba84} 
and in section 3 we 
examine the solutions which satisfy the correct boundary conditions, 
i.e., of scattering wave type. In section 4 these considerations 
are illustrated 
for the particular case of a square well potential and a constant 
pairing field of a finite range. Then we discuss how the resonant
continuum can be taken into account in  HF-BCS approach.
Finally we show 
how the continuum, calculated in different approximations, can
affect the pairing properties of a nucleus close to the neutron drip line.

\section{Hartree-Fock-Bogoliubov equations in coordinate space}

Let us denote by $H$ the hamiltonian of the nuclear system. 
When treated in the HFB approximation, $H$ will give rise to the mean 
field and the pairing field in which 
the quasiparticles are moving. We assume that the two-body effective interactions 
associated with the Hartree-Fock (HF) mean field and 
pairing field are zero-range forces so that the 
total energy $E$ is a functional of the {\it local} particle density $\rho({\bf r})$ 
and pairing density $\kappa({\bf r})$: 
\begin{eqnarray}
\rho({\bf r}) & \equiv & \langle HFB \vert \Psi^+({\bf r}) \Psi({\bf r}) \vert 
HFB \rangle~, \nonumber \\
\kappa({\bf r}) & \equiv & \langle HFB \vert \Psi^+({\bf r}) \Psi^+({\bf r}) 
\vert HFB \rangle~, 
\label{eq1}
\end{eqnarray}
where $\Psi^+({\bf r})$ is a nucleon creation operator and $\vert HFB \rangle$ 
is the HFB ground state. Typically, the self-consistent mean field is generated by a 
Skyrme-type interaction whereas the pairing field is produced by a zero-range, 
possibly density-dependent interaction.  

The HFB equations in coordinate space can be expressed in the following form:
\begin{eqnarray}
\pmatrix{h - \lambda & \Delta \nonumber \\ \Delta & - (h - \lambda)} 
\pmatrix{ U_\alpha ({\bf r}) \nonumber \\ V_\alpha ({\bf r})} = E_\alpha 
\pmatrix{ U_\alpha ({\bf r}) \nonumber \\ V_\alpha ({\bf r})}~, 
\label{eq2}
\end{eqnarray}
where $\lambda$ is the chemical potential, $h$ is the HF hamiltonian 
and $\Delta$ is the pairing field. 
The local nucleon density and pairing density can be written in terms of the solutions 
$(U_\alpha , V_\alpha)$ in the form:
\begin{eqnarray}
\rho({\bf r}) & = & \sum_\alpha \vert V_\alpha ({\bf r})\vert^2~,
\label{eq3}
\end{eqnarray}
\begin{eqnarray}
\kappa({\bf r}) & = & \sum_\alpha U_\alpha ({\bf r}) V^{*}_\alpha ({\bf r})~.
\label{eq4}
\end{eqnarray}
For Skyrme forces, 
the HF and pairing fields may also depend on derivatives of $\rho$ and
$\kappa$ and on some local currents  
because of the velocity dependence of the 
interaction. Since all densities appearing in the energy functional are local, 
Eq.(\ref{eq2}) is a set of coupled differential 
equations which is highly non-linear because the fields $h$ and $\Delta$ depend 
themselves on the solutions ${ U_\alpha, V_\alpha }$ (self-consistency problem). 
Nevertheless, using Skyrme-type forces brings a major simplification. If one starts 
instead with a finite range effective force, for instance a Gogny's force, 
Eq.(\ref{eq2}) would be a set of coupled integro-differential equations which is 
usually solved in a harmonic oscillator basis. 
The general properties of Skyrme-HF hamiltonians
are well-known and 
detailed expressions of 
$h$ and $\Delta$ for the case of Skyrme forces can be found for instance in 
Ref.~\cite{Doba84}. 

The Skyrme-HF hamiltonian has the general form:
\begin{eqnarray}
h & = & -{\bf \nabla}.\frac{\hbar^2}{2m^*({\bf r})}{\bf \nabla} + V_{HF}({\bf r}), 
\label{eq5}
\end{eqnarray}
where the effective mass $m^*$ and HF potential
$V_{HF}$ depend on nucleon densities
and currents. An important property is that $m^*$
tends to the nucleon mass $m$ and 
$V_{HF}$ tends to zero at infinity, with the same rate as the 
densities go to zero. We shall use this property in the next section to 
establish the asymptotic forms of the solutions of Eq.(\ref{eq2}).

For illustration, we can look at the pairing field calculated with a 
pairing interaction often used in the literature~\cite{Doba96,Tera96}:
\begin{eqnarray}
V(1,2) & = & V_0 \left[ 1 - \left( \frac {\rho({\bf r}_1)}
{\rho_c}\right)^{\gamma}~\right]  \delta({\bf r}_1 - {\bf r}_2)~.
\label{eq6}
\end{eqnarray}
To be meaningful, such an interaction must be used with an energy
cut-off in the
quasiparticle spectrum, i.e., the summations in Eqs.(\ref{eq3},\ref{eq4}) 
must be
limited to $E_\alpha \le E_{cutoff}$. 
In this case, the pairing field is:
\begin{eqnarray}
\Delta({\bf r}) & = & \frac {1}{2} V_0 \left[ 1 - \left( \frac {\rho({\bf r})}
{\rho_c}\right)^{\gamma}~\right] \kappa({\bf r})~. 
\label{eq7}
\end{eqnarray}
On this example we can see that the local pairing field behaves asymptotically 
like the local pairing density.

\section{Asymptotic behaviour of HFB solutions}

 From the symmetries of Eq.(\ref{eq2}) it can be seen that, if
$(E_\alpha, U_\alpha, V_\alpha)$ is a solution then
$(-E_\alpha, V^*_\alpha, U^*_\alpha)$ is also a solution. Thus, we need to
consider only one class of solutions and we choose those with $E_\alpha \ge
0$. 

The asymptotic behaviour of the solutions of Eq.(\ref{eq2}) has been discussed 
in detail by Bulgac~\cite{Bulgac}. Let us consider for simplicity the case of
spherical symmetry and write:
\begin{eqnarray}
\pmatrix{ U_\alpha ({\bf r}) \nonumber \\ V_\alpha ({\bf r})} = \frac 
{1}{r}
\pmatrix{ u_\alpha ({r}) \nonumber \\ v_\alpha ({r})} 
 {\bf Y}_\alpha (\hat r, \sigma)~. 
\label{eq8}
\end{eqnarray}
Eq.(\ref{eq2}) now becomes a one-dimensional equation  in the radial
coordinate $r$ for each $\alpha = (l,j)$. At very large distances the HF
hamiltonian $h$ tends to $-(\hbar^2/2m)(\frac{1}{r}\frac{d^2}{dr^2}r -
\frac{l(l+1)}{r^2})$ (plus a Coulomb potential for protons) while the pairing
field $\Delta(r)$ has vanished. The equations for
$u_\alpha ({r})$ and $v_\alpha ({r})$ are decoupled and one can easily see
how the physical solutions must behave at infinity.
Thus, for a negative
chemical potential $\lambda$, i.e., for a
bound system, there are two well separated
regions in the quasiparticle spectrum:\\
- between 0 and $-\lambda$ the quasiparticle spectrum is discrete
  and both upper and lower components
 $ (u_\alpha ({r}), v_\alpha ({r}))$ of the HFB wave function 
 decay   exponentially at infinity;\\ 
- above $-\lambda$ the quasiparticle spectrum is continuous and the physical
  solutions are such that at infinity the upper component of the HFB wave 
  function has a scattering wave form (see the next section) while the 
  lower component is exponentially decaying. In what follows the
  continuous HFB wave functions are normalised to a delta function 
  of energy. 

Thus, the summations in Eqs.(\ref{eq3}-\ref{eq4}) 
should in fact include integrations over the continuum of the 
quasiparticle spectrum:
\begin{eqnarray}
\rho({\bf r}) & = & \sum_{0 \le E_\alpha \le -\lambda} 
\vert V_\alpha({\bf r})\vert^2 + \int_{-\lambda}^{E_{cutoff}} dE_\alpha 
\vert V_{E_\alpha}({\bf r})\vert^2~, \nonumber \\ 
\kappa({\bf r}) & = & \sum_{0 \le E_\alpha \le -\lambda}U_\alpha({\bf r}) 
V^{*}_\alpha({\bf r}) + \int_{-\lambda}^{E_{cutoff}} dE_\alpha 
U_{E_\alpha}({\bf r}) V^{*}_{E_\alpha}({\bf r})~. 
\label{eq9}
\end{eqnarray}

As we have already mentioned in the introduction, most of the practical 
calculations are done by solving Eq.(\ref{eq1}) with a box boundary 
condition~\cite{Doba84,Doba96,Tera96}, 
i.e., by requiring that the solutions 
$U_\alpha({\bf r}), V_\alpha({\bf r})$ 
have a node at $r = R$ 
(they are taken to be identically zero beyond $R$). This condition makes the 
spectrum of $E_\alpha$ entirely discrete and allows the use of 
Eqs.(\ref{eq3}-\ref{eq4}) instead of Eq.(\ref{eq9}). However, it is not clear 
how accurately one can mock up continuum effects like single-particle 
resonance contributions by using this discretization procedure. 
In order to avoid such ambiguities one needs solutions with proper
boundary conditions. These solutions are illustrated in the next section 
for the case of a simple model~\cite{Bulgac,Belyaev}.

\section{HFB solutions: a schematic model}

 In what follows we discuss the solutions of HFB equations in coordinate
space for a mean field given by a square well potential, of depth $V_0$
and radius $a$, and a pairing field, $\Delta$, constant inside the same
radius $a$  
and zero outside. We suppose also that the Fermi
level, $\lambda$, is given. For such a system the radial HFB  equations 
inside the potential well, i.e. for $r\le a$, are given by:
\begin{eqnarray}
 (\frac{1}{r}\frac{d^2}{dr^2}r - \frac{l(l+1)}{r^2}+\alpha^2) u_{lj}
-\delta^2 v_{lj}& = & 0~, 
\nonumber \\ 
 ( \frac {1}{r} 
\frac {d^2}{dr^2} r 
- \frac{l(l+1)}{r^2}+\beta^2) v_{lj}
-\delta^2 u_{lj} & = & 0~.
\label{eq10}
\end{eqnarray}
where $\alpha^2=\frac{2m}{\hbar^2}(\lambda+E+U_0)$,
$\beta^2=\frac{2m}{\hbar^2}(\lambda-E+U_0)$, 
$\delta^2=\frac{2m}{\hbar^2}\Delta$ and $U_0=-(V_0+V_{so}\vec l. \vec s)$.
 The above equations have the following physical solution for any value 
 of the quasiparticle energy :
\begin{eqnarray}
u_{lj} & = & A_{+}j_l(k_+r) + A_{-}j_l(k_{-}r)~, \nonumber \\ 
v_{lj} & = & A_{+}g_{+}j_l(k_+r) + A_{-}g_{-}j_l(k_{-}r)~,
\label{eq11}
\end{eqnarray}
where $j_l$ are spherical Bessel functions, 
$k_{\pm}=\frac{2m}{\hbar^2}(U_0+\lambda \pm
\sqrt(E^2-\Delta^2))$ and 
$g_{\pm}=(E \pm \sqrt(E^2-\Delta^2))/\Delta$.

Outside the potential well both $U_0$ and $\Delta$ are zero and the 
HFB equations are decoupled. In this case the type of solutions 
depends on the quasiparticle energy. 
Thus, for $E < -\lambda$ the solutions have the form:
\begin{eqnarray}
u_{lj} & = & A h_l^{(+)}(\alpha_1r)~, \nonumber \\ 
v_{lj} & = & B h_l^{(+)}(\beta_1r)~,
\label{eq12}
\end{eqnarray}
where $h_l^{(+)}$ are spherical Haenkel functions, 
$\alpha_1^2=\frac{2m}{\hbar^2}(\lambda+E)$ and
$\beta_1^2=\frac{2m}{\hbar^2}(\lambda-E)$. 
These solutions correspond to the bound quasiparticle spectrum.

For $E > - \lambda$ the spectrum is continuous and the solutions are:
\begin{eqnarray}
u_{lj} & = & C [cos(\delta_{lj}) j_l(\alpha_1r)-sin(\delta_{lj})
n_l(\alpha_1r)]~, \nonumber \\ 
v_{lj} & = & D h_l^{(+)}(\beta_1r)~,
\label{eq13}
\end{eqnarray}
where $n_l$ are spherical Neumann functions and $\delta_{lj}$ 
is the phase shift corresponding to the angular momentum $(lj)$.

 The constants entering in the wave functions above 
 are fixed by the continuity conditions of the solutions and of their
 derivatives  at $r=a$ and the normalisation.
 In what follows we discuss only the continuous solutions, i.e., for 
 energies $E > -\lambda$.
 
 Of particular interest are the values of energies for which 
 the wave functions have maximum localisations inside the potential well. 
 These are the regions close to quasiparticle resonances, which can be
 defined as complex outgoing solutions of HFB equations.
 In HFB approach one distinguishes two types of quasiparticle 
 resonances. One  type corresponds to the single-particle 
 resonances of the mean field. For the pairing correlations an important
 role is played by those single-particle resonances which are close to 
 the particle thereshold and have relatively high angular momentum.

 Another type of quasiparticle resonances corresponds to bound 
 single-particle states. 
 These resonant states, which appear due to the non-diagonal matrix 
 elements of the pairing field, are specific of the HFB approach. 
 The resonant states corresponding to deep hole states have 
 small widths. The states with very small widths
 can be eventually treated as quasibound states, normalized to unity in 
 the same volume as the bound quasiparticle states. 
  
  A special attention is paid usually to the continuum s1/2 states 
  \cite{Belyaev,Bennaceur}. Apart from
  the deep hole s1/2 states, which in HFB become  narrow 
  quasiparticle resonances with high quasiparticle energies, 
  in drip line nuclei one may also find a loosely bound s1/2 
  single-particle state. In continuum HFB approach this state 
  appears often as a broad quasiparticle resonance (see example
  below), close to the continuum edge, and  its role in pairing 
  correlations cannot be distinguished from the rest of non-resonant 
  s1/2 continuum.
  This is different from a HF-BCS approach, where the contribution
  of the loosely bound single-particle s1/2 state to pairing correlations 
  is not mixed with the rest of background single-particle continuum.

  The structure of the continuum discussed above is essentially 
  the same for general, self-consistent HFB calculations. In this case
  the HFB equations for the continuum spectrum are integrated by starting
  far from the nucleus with the solution given by Eq.(\ref{eq13}), which is 
  propagated towards the matching point by the Numerov method. For 
  each quasiparticle energy one calculates, by matching conditions, 
  the phase shift. Then the energies
  (widths) of quasiparticle resonances are found from the energies where the
  derivative of the phase shift reaches its maximum (half of its maximum)
  value. This information is used afterwards to fix an appropriate
  energy grid (i.e., dense in the energy region of a resonance) for the
  calculation of the continuum contribution to the particle and pairing 
  densities (see Eqs.(\ref{eq9})).

\section{HF-BCS approximation}

 The HF-BCS approximation is obtained by neglecting in the HFB equations
 the non-diagonal matrix elements of the pairing field. This means that 
 in the HF-BCS limit one neglects the pairing correlations induced by the 
 pairs formed in states which are not time-reversed partners.  
 
 Particularly simple are the HF-BCS equations which include the
 effect of resonant continuum \cite{Saliwy,Sagili}:
\begin{eqnarray}
\Delta_i & = & \sum_j \langle i,\bar i \vert V \vert j, \bar j \rangle u_j
v_j + \sum_{\nu}  
\langle i,\bar i \vert V \vert \nu \epsilon_\nu, \overline{\nu \epsilon_\nu} 
\rangle
\int_{I_\nu} g^{c}_\nu(\epsilon) 
 u_{\nu}(\epsilon) v_{\nu}(\epsilon) d\epsilon~, \nonumber \\
\Delta_{\nu}(\epsilon) &  = & (g^{c}_{\nu}(\epsilon)/g_{\nu}(\epsilon)) \big(
\sum_j \langle \nu \epsilon, \overline{\nu
\epsilon} \vert V \vert j, \bar j \rangle u_j
v_j \nonumber \\ 
 & + & \sum_{\nu \prime} \langle \nu \epsilon_\nu,\overline{\nu \epsilon_\nu} 
 \vert V \vert 
\nu \prime \epsilon_{\nu \prime}, \overline{\nu \prime \epsilon_{\nu \prime}} 
\rangle 
 \int_{I_\nu \prime} g^{c}_{\nu \prime}(\epsilon \prime) 
 u_{\nu \prime}(\epsilon \prime) v_{\nu \prime}(\epsilon \prime) 
d\epsilon \prime 
\big) \nonumber \\
 & \equiv & (g^{c}_{\nu}(\epsilon)/g_{\nu}(\epsilon)) \Delta_\nu~.
\label{eq14}
\end{eqnarray}
where
$g^c_\nu(\epsilon) = \frac {2j_\nu + 1}{\pi} \frac{d\delta_\nu}{d\epsilon}$, 
$g_\nu(\epsilon)$ is the total level density and
$\delta_\nu$ is the phase shift of angular momentum $(l_{\nu} j_{\nu})$.
In these equations the interaction matrix elements are  calculated with
the scattering wave functions at  resonance energies and normalised
inside the volume where the pairing interaction is active. 
The particle number condition is:
\begin{eqnarray}
N = \sum_i v_i^2 + \sum_\nu \int_{I_\nu} g^c_{\nu}(\epsilon) v^2_\nu
(\epsilon) d\epsilon~.
\label{eq15}
\end{eqnarray}
The energy factor $g^c_\nu(\epsilon)$ takes into account the variation
of the localisation of scattering states in the energy region of
a resonance ( i.e., the width effect) and goes  to a delta function in the 
limit of a very narrow width. For more details one can see 
Ref.\cite{Sagili}. 

\section{Application: continuum coupling in different approximations}

 In order to illustrate the approximations discussed above
 we take as an example the nucleus
 $^{84}$Ni, for which HFB calculations with discretized continuum 
 can be found in the literature \cite{Tera96}.

\begin{table}[h]
\label{table1}
\caption{Energies and widths of quasiparticle resonant states. 
In the two 
last columns the total occupancies 
for each (lj) channel are shown 
both in full and in box calculations. }

\vspace{0.2cm}
\begin{center}
\footnotesize
\begin{tabular}{|cc|cc|cc|cc|cc|c|}
\hline
j & & l & & E(MeV) & & 
$\Gamma$(MeV) & &
Total occ. (full) & & Total occ. (box)  \\
\hline
 1/2 & & 0  & & 1.241  & & 0.69  & &   & &   \\
 1/2 & & 0  & & 20.956  & & 0.154  & &   & &   \\
 1/2 & & 0  & & 43.488  & & 10$^{-6}$  & &   & &   \\
  & &   & &   & &   & & 2.261  & & 2.299  \\
 1/2 & & 1  & & 8.076  & & 0.36  & &   & &   \\
 1/2 & & 1  & & 33.531  & & 0.101  & &   & &   \\
  & &   & &   & &   & & 1.993  & & 1.994  \\
 3/2 & & 1  & & 9.806  & &  0.60 & &   & &   \\
 3/2 & & 1  & & 35.058  & & 0.104  & &   & &   \\
  & &   & &   & &   & & 1.999  & & 2.016  \\
 3/2 & & 2  & & 2.354  & & 0.641  & &   & &   \\
 3/2 & & 2  & & 22.103  & & 0.066  & &   & &   \\
  & &   & &   & &   & &  1.171 & & 1.183  \\
 5/2 & & 2  & & 1.816  & & 0.076  & &   & &   \\
 5/2 & & 2  & & 25.713  & & 0.006  & &   & &   \\
  & &   & &   & &   & &  1.587 & & 1.529  \\
 5/2 & & 3  & & 8.972  & & 0.919  & &   & &   \\
  & &   & &   & &   & & 0.997  & & 1.006  \\
 7/2 & & 3  & & 15.449  & & 2.070  & &   & &   \\
  & &   & &   & &   & & 1.011  & &  1.015 \\
 7/2 & & 4  & & 3.492  & & 0.027  & &   & &   \\
  & &   & &   & &   & & 0.101  & & 0.110  \\
 9/2 & & 4  & &  5.747 & & 0.001  & &   & &   \\
  & &   & &   & &   & & 0.958  & & 0.952  \\
 11/2 & & 5  & & 5.275  & & 0.049  & &   & &   \\
  & &   & &   & &   & & 0.068  & & 0.079  \\
\hline
\end{tabular}
\end{center}
\end{table}

In all calculations discussed below the  HF mean field is calculated with 
the Skyrme interaction SIII. The pairing force is given by
Eq.(\ref{eq6}) and the parameters are the ones used in Ref ~\cite{Tera96}.

 First, we present the  results given by the continuum HFB equations
 solved with proper boundary conditions, as defined in the previous
 sections. These results, referred below as "full" continuum 
 calculations, are compared to the HFB calculations performed with 
 box boundary conditions. 

In Table 1 are shown the quasiparticle
 resonant states, of hole and particle type. For each $(lj)$ channel
 are shown also the total occupancy obtained in full and box calculations.
 One can notice the large widths of quasiparticle states corresponding
 to bound single-particle states close to the Fermi energy.
The widths are obtained from the phaseshift behaviour around $\pi/2$ except
for the three states $s1/2$, $p1/2$ and $p3/2$ close to the Fermi energy for
which the widths are extracted from the occupancy profiles. 

 The results for the pairing field, pairing density and particle density 
 are shown in Figs.(1-3). On can see that the box calculations generally 
 overestimate the pairing correlations. 
 This is also seen by comparing 
 the Fermi levels (full: -1.026 MeV; box: -1.205 MeV), averaged gaps 
 (full: 1.230 MeV; box: 1.448 MeV), pairing energies (full: -18.092 MeV; 
 box: -24.081 MeV) and total binding energies (full: -651.079 MeV; 
 box:-654.708 MeV). 

Next, we compare the results given by the HF-BCS approximation presented 
in the previous section, to the HFB calculations of Ref.\cite{Tera96}. 
In Ref. \cite{Tera96} the HFB equations are diagonalised in a basis 
formed by the single-particle states selected by a  box of a finite radius. 
The single-particle continuum is cut at 5 MeV. This energy cutoff is 
different from the HFB box calculations presented above, where the energy
cutoff is much higher. Up to 5 MeV, one finds three single-particle 
resonances, $d_{3/2}$, $g_{7/2}$ and $h_{11/2}$, which are treated in 
the HF-BCS approximation given by Eqs.(\ref{eq14}).
The HFB and HF-BCS results for the pairing field are shown in Fig.4. 
In the same figure is shown  the HF-BCS result obtained by replacing 
each single-particle resonance by a  discrete state, 
normalized in the same volume as the one used for the bound states.
As seen from Fig.4, this result  is closer to HFB calculations. This 
similarity is due to the fact that in the HFB calculations each
resonant state is actually represented by a unique state normalised
in the box. When the effect of the width of resonant states is 
included in HF-BCS equations, the pairing correlations are decreasing.
This trend is similar with the one observed in continuum HFB calculations
presented above. Thus, one can conclude that, 
in order to take fully into account 
the effect of resonant continuum upon pairing correlations, one needs to 
solve the HF-BCS and HFB equations with proper boundary conditions.


\begin{figure}[ht]
\begin{center}
\includegraphics*[scale=0.45,angle=-90.]{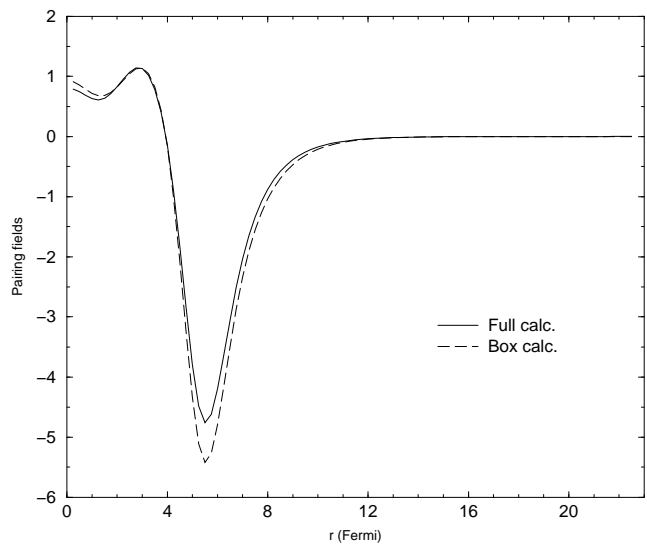}
\caption{Pairing fields in full (full line) and box (dashed line) 
calculations. }
\label{fig1}
\end{center}
\end{figure}

\begin{figure}[hb]
\begin{center}
\includegraphics*[scale=0.5,angle=-90.]{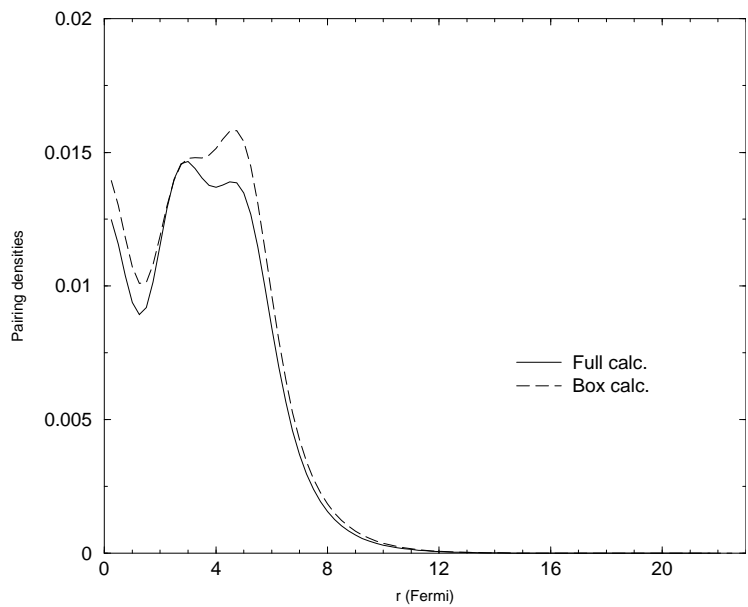}
\caption{Pairing densities in full (full line) and box (dashed line) 
calculations. }
\label{fig2}
\end{center}
\end{figure}

\newpage

\begin{figure}[ht]
\begin{center}
\includegraphics*[scale=0.40,angle=-90.]{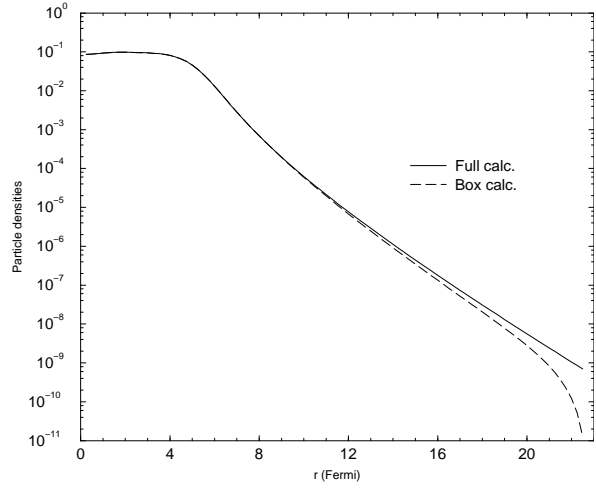}
\caption{Particle densities in full (full line) and box (dashed line) 
calculations. }
\label{fig3}
\end{center}
\end{figure}

\begin{figure}[hb]
\begin{center}
\includegraphics*[scale=0.40,angle=-90.]{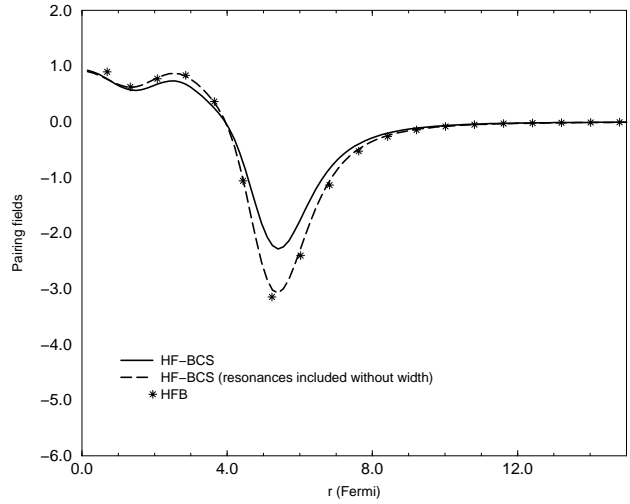}
\caption{Pairing fields calculated in different approximations: HFB of Ref.
  \cite{Tera96} (stars), resonant HF-BCS with widths (solid) and without
  widths (dashed). }
\label{fig4}
\end{center}
\end{figure}

\newpage

\section{Summary}

 In these lectures, we have discussed how the continuum coupling
 and pairing correlations are calculated in drip line nuclei.
 The discussion was restricted to the HFB and HF-BCS approximations.
 In the first part, we have shown how one can construct in the
 HFB approach solutions with proper boundary conditions for the 
 continuum spectrum. Then, for the case of a simple model, we have
 analysed the structure of quasiparticle continuum. In particular 
 we have discussed the treatment of resonant continuum both in
 HFB and HF-BCS approaches.

 In the second part we have shown, for the case of a neutron 
 rich nucleus, how different treatments of continuum can affect 
 the pairing correlations. One concludes that the solutions based 
 on discretized continuum (box boundary conditions) can overestimate
 the pairing correlations when compared with proper continuum HFB 
 or HF-BCS calculations. This may have consequences on the predictions of
 drip lines.

\begin{acknowledgments}
We thank J. Dobaczewski for providing us the code which
solves the HFB equations with box boundary conditions.
\end{acknowledgments}



\begin{chapthebibliography}{1}
\bibitem{Ring-Schuck}
P. Ring and P. Schuck, {\em The Nuclear Many-Body Problem} 
(Springer-Verlag, New York, 1990). 

\bibitem{Doba84}
J. Dobaczewski, H. Flocard and J. Treiner, {\it Nucl. Phys.} 
{\bf A 422}, 103 (1984).

\bibitem{Doba96}
J. Dobaczewski, W. Nazarewicz, T.R. Werner, J.F. Berger, C.R. Chinn and 
J. Decharg\'e, {\it Phys. Rev.} {\bf C 53}, 2809 (1996).


\bibitem{Tera96}
J. Terasaki, P.-H. Heenen, H. Flocard and P. Bonche, {\it Nucl. Phys.} 
{\bf A 600}, 371 (1996).

\bibitem{Bulgac}
A. Bulgac, preprint nucl-th/9907088.

\bibitem{Belyaev}
S.T. Belyaev, A.V. Smirnov, S.V. Tolokonnikov and S.A. Fayans, 
{\em Sov. J. Nucl. Phys.} {\bf 45}, 783 (1987).

\bibitem{Bennaceur}
K. Bennaceur, J. Dobaczewski and M. Ploszajjczak, 
{\it Phys. Rev.} {\bf C 60}, 2809 (1999).

\bibitem{Saliwy}
N. Sandulescu, R.J. Liotta and R. Wyss, {\it Phys. Lett.} 
{\bf B 394}, 6 (1997).

\bibitem{Sagili}
N. Sandulescu, N. Van Giai and R.J. Liotta, {\it Phys. Rev.} {\bf C 61}
(2000) 061301 (R).

\end{chapthebibliography}
\end{document}